\documentclass[aps,prb,reprint,superscriptaddress,floatfix]{revtex4-2}

\usepackage[T1]{fontenc}
\usepackage[utf8]{inputenc}
\usepackage{lmodern}
\usepackage{amsmath,amssymb,bm}
\usepackage{graphicx}
\usepackage{booktabs}
\usepackage{siunitx}
\usepackage{xspace}
\usepackage{xcolor}
\usepackage{comment}

\usepackage{xcolor}

\graphicspath{{.}}

\newcommand{\YIG}{Y$_3$Fe$_5$O$_{12}$\xspace}

\begin{document}

% ============================================================
\title{Phonon-driven tuning of exchange interactions in Y$_3$Fe$_5$O$_{12}$}

\author{Kunihiko Yamauchi}
\author{Tamio Oguchi}
\affiliation{Center for Spintronics Research Network, The University of Osaka, Toyonaka, Osaka 560-8531, Japan}

\date{\today}

% ============================================================
\begin{abstract}
Yttrium iron garnet (Y$_3$Fe$_5$O$_{12}$) is a prototypical ferrimagnetic insulator widely used in spin-wave and magnonic devices owing to its extremely low magnetic damping and long magnon propagation length, and recent experiments suggest that lattice vibrations can influence magnetic properties, motivating a microscopic understanding of how phonons modify exchange interactions. In this work, phonon-driven tuning of exchange interactions in Y$_3$Fe$_5$O$_{12}$ is investigated from a mode-resolved perspective based on first-principles calculations. We focus on how optical phonons modify the dominant superexchange pathways and how lattice distortions affect the Fe--O--Fe bond geometry that governs the exchange interaction. To this end, phonon modes are computed from density functional theory, and the exchange interactions are evaluated from a Wannier-based tight-binding model and mapped onto a spin Hamiltonian, while displaced structures along individual infrared-active modes are used to quantify their impact on the magnetic interactions. 
\end{abstract}
\maketitle

% ============================================================

\section{Introduction}

Magnons, the quanta of spin-wave excitations, constitute a central platform for low-dissipation information transport in magnetic insulators.
Among known materials, yttrium iron garnet (Y$_3$Fe$_5$O$_{12}$, YIG) occupies a unique position due to its extremely small Gilbert damping and long magnon mean free path, which enable coherent spin-wave propagation over macroscopic distances.\cite{Cherepanov1993} 
These properties have made YIG a cornerstone material for magnonics and spin-wave-based device concepts.
Beyond passive transport, active control of magnons is a key challenge for future spintronic applications. 
Experimental studies have revealed signatures of magnon--phonon coupling in YIG. 
Indirect evidence was first reported in the spin Seebeck effect, where anomalous features were observed and attributed to magnon--phonon coupling.\cite{Kikkawa2016PRL} 
Subsequent inelastic neutron scattering measurements directly confirmed this picture by observing hybridized magnon--phonon excitations at selected wave vectors, demonstrating the formation of magnon--polaron states.\cite{Man2017}

In contrast to spin--orbit-driven systems such as CrI$_3$, where strong anisotropy leads to pronounced magnon--phonon hybridization,\cite{WebsterYan2018_spinphonon_cri3,Delugas2023PRB,Fang2025}
YIG belongs to a weak spin--orbit regime in which lattice vibrations primarily modulate the exchange interactions.
In YIG, the magnon dispersion has been well established experimentally over a wide energy range by inelastic neutron scattering.\cite{Plant1977}
The magnon--phonon coupling has been studied using both model and first-principles approaches. 
Flebus \textit{et al.} proposed a model description of magnon--polaron transport, and showed that exchange-striction can produce avoided-crossing-like features even without forming well-defined hybrid quasiparticles.\cite{Flebus2017PRB} 
In parallel, first-principles studies have investigated the impact of lattice vibrations on magnon properties in YIG.\cite{Liu2017,Wang2020}
From this perspective, it is of interest to quantify how individual phonon modes modify the magnetic exchange interactions and thereby influence the magnon spectrum.
At the same time, certain optical phonon modes carry large mode effective charges and can be efficiently driven by external electric fields.\cite{KingSmith1993,Resta1994} 
If such modes strongly modulate the dominant exchange interactions, they provide a route to electric-field control of magnons even in centrosymmetric materials such as YIG.\cite{Baettig2008}

In this work, we examine the following two aspects on the basis of first-principles calculations.
First, we quantitatively evaluate the exchange-striction effect in YIG by combining calculations of electronic structure, phonons, and magnetic exchange interactions.
Second, we identify infrared-active phonon modes that produce large modulation of exchange interactions and discuss their relevance for electric-field control of magnetic interactions.

\section{Crystal structure and computational method}

\begin{figure}[t]
  \centering
  \includegraphics[width=0.8\columnwidth]{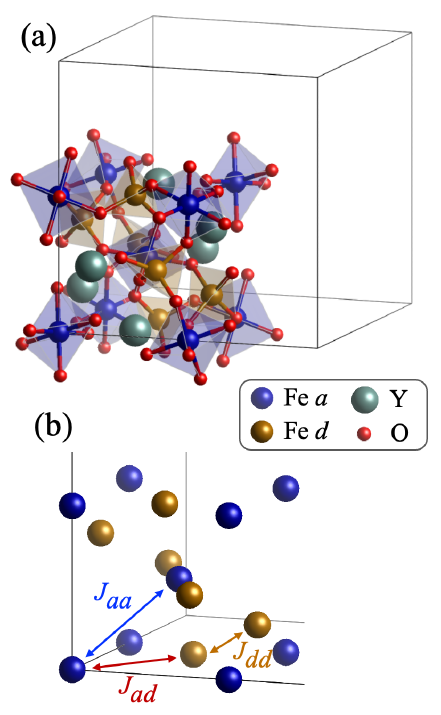}
  %{structure.pdf}
\caption{
  Crystal structure of Y$_3$Fe$_5$O$_{12}$.
  Octahedral $a$ sites and tetrahedral $d$ sites form two magnetic sublattices.
  The dominant $J_{ad}$ and additional $J_{dd}$ and $J_{aa}$ exchange paths are highlighted.
}
  \label{fig:structure}
\end{figure}

YIG crystallizes in a cubic garnet structure with space group $Ia\bar{3}d$ as shown in Fig.~\ref{fig:structure}.
Fe$^{3+}$ ions occupy the octahedral (16$a$) and tetrahedral (24$d$) Wyckoff positions, forming a ferrimagnetic ground state with antiparallel alignment between the two sublattices. 
The magnetic interactions in YIG are well described by a Heisenberg model with several exchange paths.
Among them, the nearest-neighbor interaction between $a$- and $d$-site Fe ions ($J_{ad}$) is dominant and sets the overall magnon energy scale.\cite{Harris1963,Xie2017}
This interaction arises from superexchange through oxygen ions and is therefore sensitive to the Fe$_a$--O--Fe$_d$ bond geometry, in accordance with the Goodenough--Kanamori rules\cite{Goodenough1955,Kanamori1959,Anderson1950,Anderson1952}.

First-principles calculations were performed within density functional theory using the Vienna \textit{ab initio} simulation package (VASP)\cite{Kresse1996,Kresse1999} within the Perdew--Burke--Ernzerhof (PBE) generalized gradient approximation\cite{Perdew1996}.
A $\Gamma$-centered $4\times4\times4$ $k$-point mesh was used for Brillouin-zone sampling.
Calculations were carried out both with and without an on-site Hubbard $U$ for Fe $3d$ states.
The lattice parameters and internal atomic coordinates were fully relaxed prior to further calculations. 

Phonon properties were obtained using the finite-displacement method as implemented in the phonopy code\cite{Togo2015}.
Born effective charges were evaluated within density-functional perturbation theory\cite{Gonze1997,Baroni2001}. 
Frozen-phonon calculations are performed using phonon eigenvectors at the $\Gamma$ point obtained from phonopy.
For each mode, atomic displacements are constructed from the corresponding eigenvectors and applied to the reference structure to generate displaced configurations with opposite signs.
The resulting changes in the magnetic interactions are found to be insensitive to the sign of the displacement and to scale linearly with the displacement amplitude, confirming that the chosen amplitude lies within the linear-response regime.
The resulting structures are used to evaluate the dependence of magnetic exchange interactions on lattice distortions.
Further details on the construction of the displacement patterns are provided in Appendix~\ref{app:phonon_alignment}. 

Magnetic exchange interactions were evaluated using TB2J\cite{He2021, Liechtenstein1987, Solovyev2021} on the basis of Wannier functions constructed with Wannier90\cite{Mostofi2008}, following the maximally localized Wannier formalism\cite{Marzari1997,Souza2001,Marzari2012}. 
The Wannier functions were constructed by projecting onto Fe and O atomic orbitals, specifically Fe $d$ and O $p$ states, in order to capture the relevant hopping processes associated with superexchange interactions. 
%\red{write what orbital states you projected}
%The magnon spectrum was computed within linear spin-wave theory based on the extracted exchange parameters.
%Details of the spin-wave formalism and numerical implementation are provided in the Appendix.
The magnon spectrum was computed within linear spin-wave theory \cite{Holstein1940, Anderson1952} based on the extracted exchange parameters, using an in-house implementation as described in Appendix~\ref{app:lswt}. 

% ============================================================
%\section{Results}

\section{Electronic structure and dependence on Hubbard $U$}

\begin{figure}[t]
  \centering
  \includegraphics[width=0.95\columnwidth]{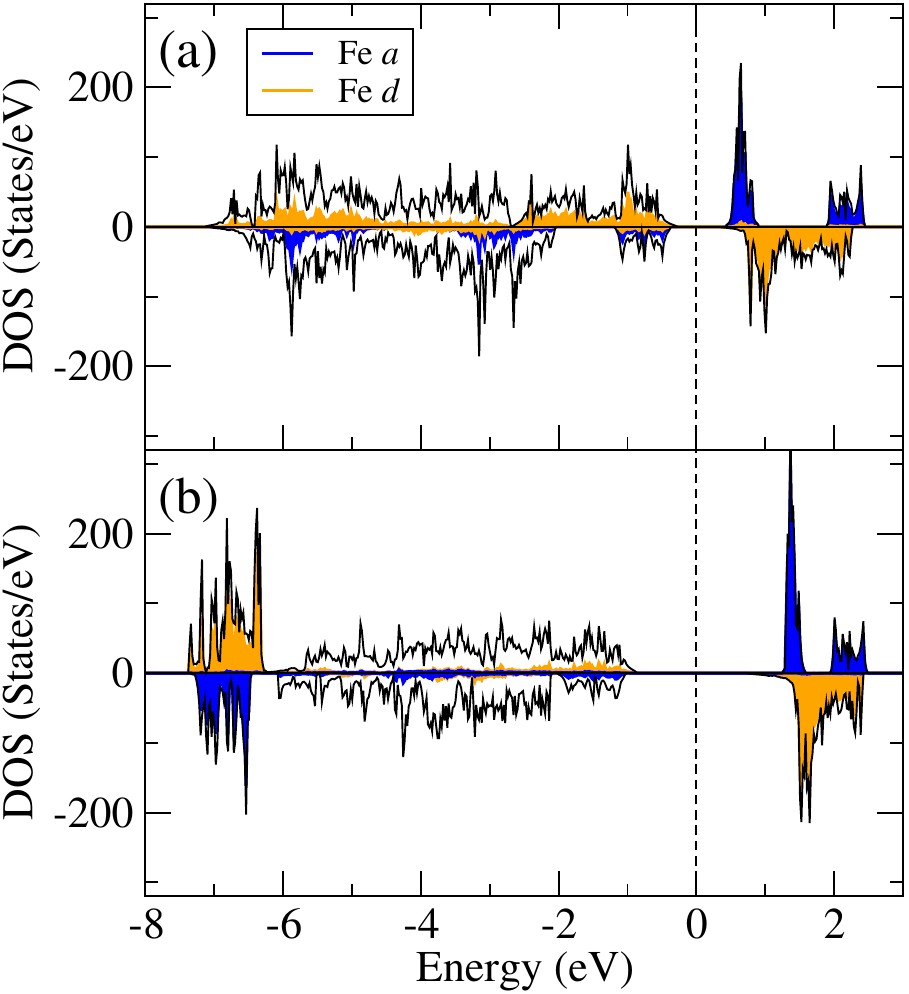} %{pdos_0u3u.pdf}
\caption{
Spin-resolved partial density of states (DOS) of \YIG\ projected onto Fe $3d$ states for (a) $U = 0$ eV and (b) $U = 3$ eV.
Contributions from Fe $a$ (octahedral, blue) and Fe $d$ (tetrahedral, orange) sites are shown together with the total density of states (black).
Increasing $U$ shifts the Fe $3d$ states to lower energies and enhances the separation between O $2p$ and Fe $3d$ states.
a}
  \label{fig:dos}
\end{figure}

\begin{figure}[t]
  \centering
  \includegraphics[width=\columnwidth]{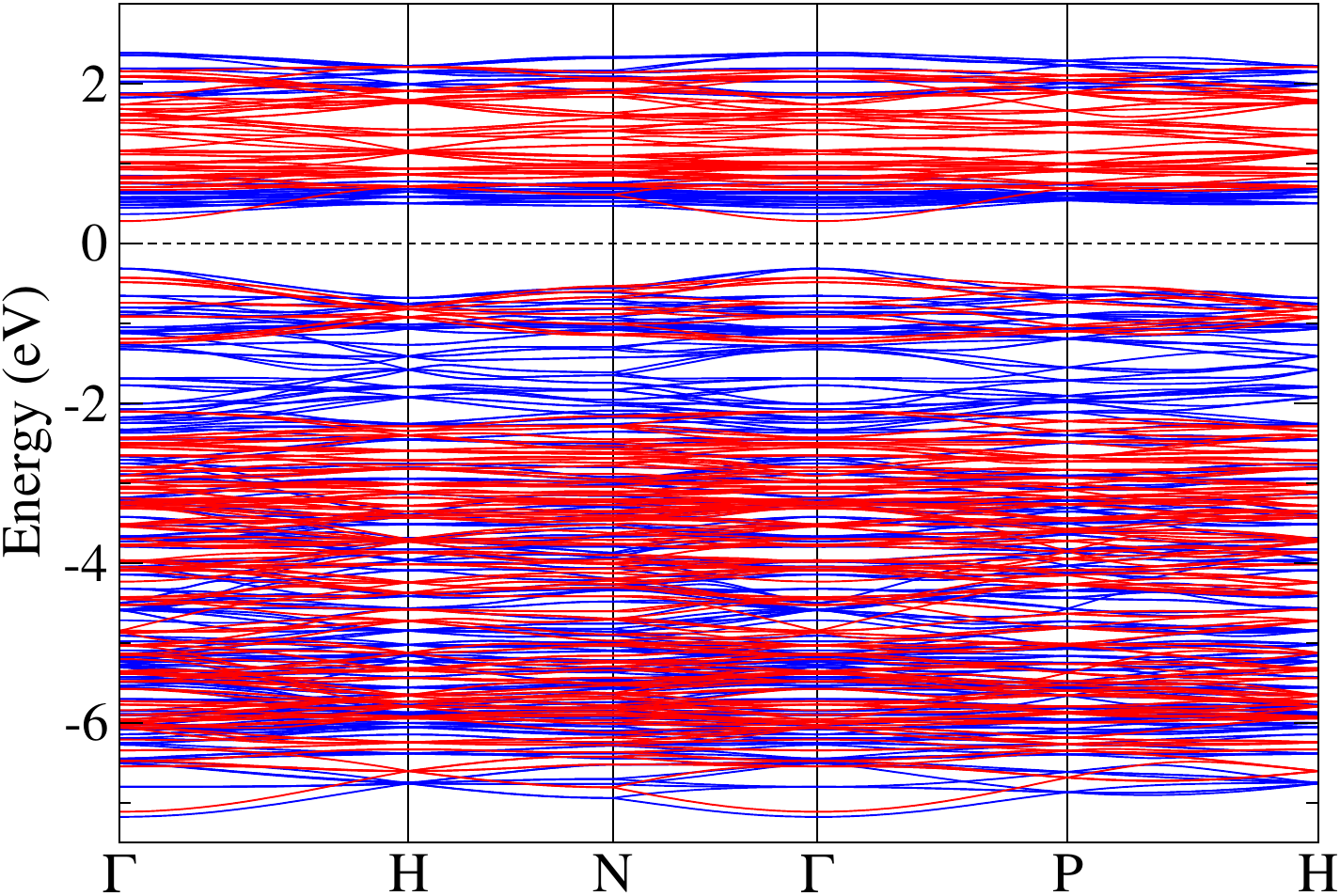} %{band_electron.pdf}
  \caption{
Electronic band structure of ferrimagnetic YIG calculated along the high-symmetry path at the GGA level ($U=0$).
Red and blue lines denote bands with opposite spin character.
}
  \label{fig:band_electron}
\end{figure}

\begin{table}[b]
\caption{
Comparison of shell-resolved exchange parameters $J_{ij}$ in Y$_3$Fe$_5$O$_{12}$ for $U=0$ and $3$ eV. 
For each pair type ($ad$, $dd$, $aa$), $J$ and $J'$ denote the nearest- and next-nearest-neighbor exchange interactions, respectively. 
The values are extracted from TB2J and given in units of meV/$S^2$ with $S=5/2$.
}
\begin{ruledtabular}
\label{tab:J1_U}
\begin{tabular}{lcccccc}
$U$ (eV) & $J^{}_{ad}$ & $J'^{}_{ad}$ & $J^{}_{dd}$ & $J'^{}_{dd}$ & $J^{}_{aa}$ & $J'^{}_{aa}$ \\
\hline
0 & -5.486 & -0.053 & -0.172 & -0.377 & \phantom{-}0.108 & -0.041 \\
3 & -4.650 & -0.027 & -0.107 & -0.307 & -0.029 & -0.007 \\
\end{tabular}
\end{ruledtabular}
\end{table}

We examine the dependence of the electronic structure and magnetic exchange interactions on the Hubbard $U$ parameter applied to the Fe $3d$ states.
Figure~\ref{fig:dos} shows that the electronic density of states is strongly affected by $U$.
At $U=0$, YIG exhibits an insulating state with mixed Fe $3d$ and O $2p$ character near the gap, placing it in an intermediate regime between Mott and charge-transfer insulators.
Increasing $U$ shifts the Fe $3d$ states to lower energies and enhances the separation between O $2p$ and Fe $3d$ states, making the system more charge-transfer-like.
On the other hand, the magnetic exchange interactions are comparatively less sensitive to $U$.
Table~\ref{tab:J1_U} summarizes the exchange parameters for $U=0$ and $3$~eV, showing only moderate changes in the dominant interactions.
While the exchange interactions remain relatively stable, the electronic structure at larger $U$ becomes less realistic.
For $U=3$~eV, the occupied Fe $3d$ states are pushed to unrealistically low energies (down to $E \lesssim E_{\mathrm{F}} - 6$~eV), which is not typical for transition-metal oxides. This reflects that the $+U$ correction shifts the occupied Fe $3d$ states to lower energies while leaving the O $2p$ states largely unchanged, thereby overestimating the O $2p$--Fe $3d$ separation. 
Based on this consideration, we adopt $U=0$ as a physically reasonable reference in the following.

Figure~\ref{fig:band_electron} shows the electronic band structure at $U=0$.
The system exhibits an insulating gap of approximately 0.6~eV, which is significantly larger than the typical energy scale of magnons.
This separation of energy scales justifies the use of an effective spin model for magnon excitations and provides a basis for the exchange-striction analysis in the following sections.

\section{Magnon dispersion and comparison with experiment}

Using the exchange parameters obtained above, we calculated the magnon dispersion of YIG.
\begin{figure}[t]
  \centering
  \includegraphics[width=0.95\linewidth]{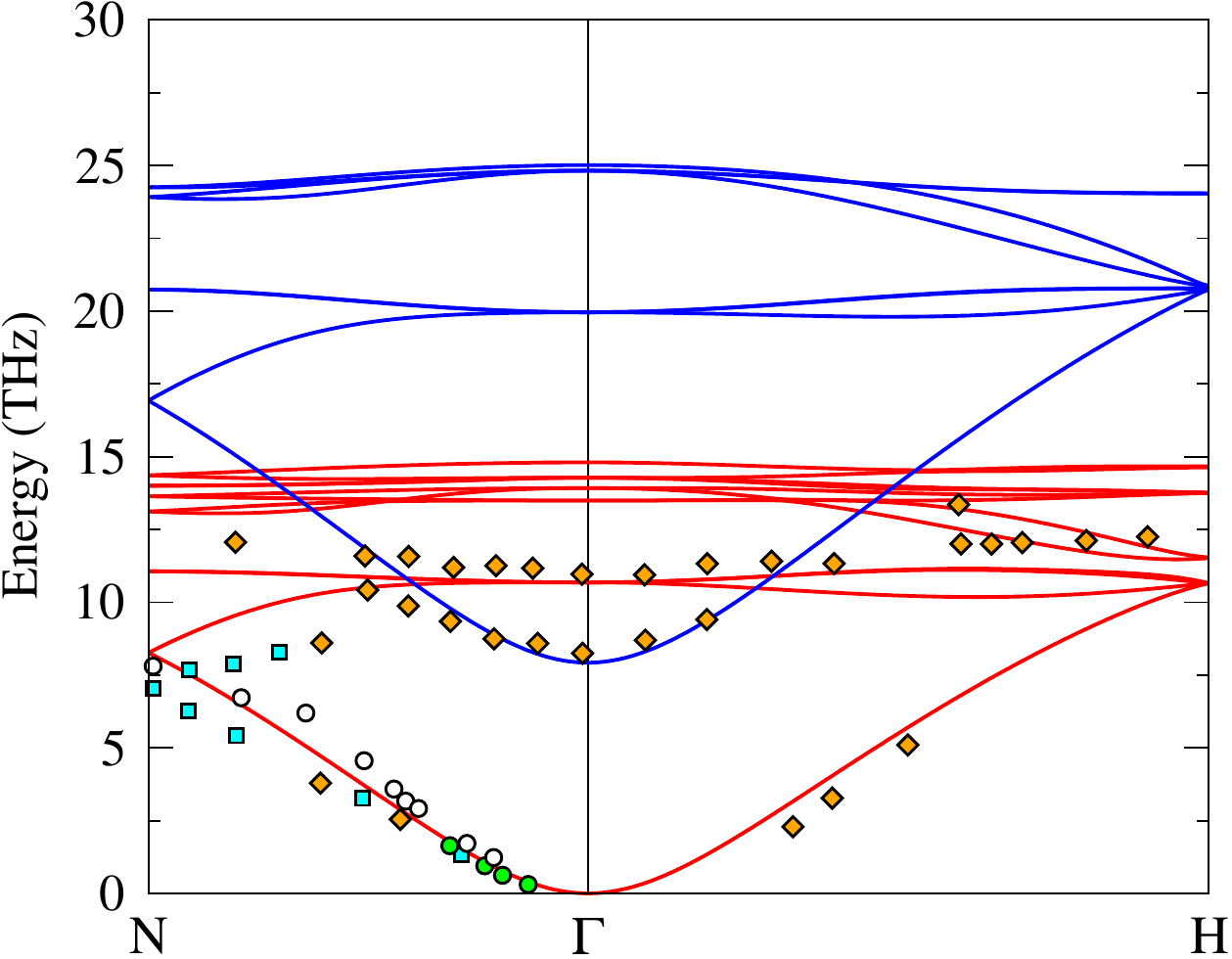} %{band_NGH_exp.pdf}
  \caption{
Calculated magnon dispersion of Y$_3$Fe$_5$O$_{12}$ with experimental data overlaid; red lines denote acoustic branches and blue lines denote optical branches.
The experimental data points are taken from the literature and were reconstructed from published figures:
The experimental data points are taken from the literature and were reconstructed from published figures:
Plant~\cite{Plant1977} (295 K, cyan filled squares; 83 K, orange filled diamonds),
Shamoto \textit{et al.}~\cite{Shamoto2018} (295 K, unfilled circles),
and Man \textit{et al.}~\cite{Man2017} (10 K, green filled circles).
}
  \label{fig:magnon}
\end{figure}
Figure~\ref{fig:magnon} compares the calculated magnon spectrum with available experimental data.
The overall energy scale and dispersion are well reproduced, including the bandwidth and the characteristic separation between low- and high-energy modes, indicating that the exchange parameters obtained from first-principles calculations provide a consistent description of the spin dynamics in YIG. 
The present results are also in good agreement with previous theoretical  studies.\cite{Gorbatov2021,Shamoto2018,Princep2017NPJ} 
The spectrum consists of 20 magnon branches per wave vector, which are naturally separated into 12 lower-energy acoustic branches and 8 higher-energy optical branches, following the established classification.\cite{Princep2017NPJ,Cherepanov1993}
This separation reflects predominantly in-phase and out-of-phase precessional motion of spins on the tetrahedral and octahedral Fe sublattices, respectively.

In the long-wavelength region near the $\Gamma$ point, the low-energy dispersion is accurately reproduced, allowing a reliable evaluation of the spin-wave stiffness defined through the quadratic dispersion $E(\mathbf{k}) = D |\mathbf{k}|^2$.
The extracted values are $D_{[100]} \approx 98$, $D_{[110]} \approx 101$, and $D_{[111]} \approx 103$ (in units of $10^{-41}$ J m$^{2}$), showing a weak but systematic directional dependence.
These values fall within the experimental range of 83--109 reported in the literature,\cite{Srivastava1987} and are close to the value obtained from QSGW calculations ($D \sim 99$), while being significantly smaller than the LDA result ($D \sim 152$).\cite{Barker2020}
In contrast to previous studies, where a single averaged value is typically reported, the present results reveal a small but finite anisotropy of the spin-wave stiffness, reflecting the directional dependence of exchange interactions in the cubic lattice. 

The overall bandwidth of the magnon dispersion is primarily governed by the nearest-neighbor interaction $J_{ad}$, while the smaller interactions $J_{dd}$ and $J_{aa}$ contribute to finer features such as splittings near the Brillouin-zone boundaries.
The present set of exchange parameters is therefore sufficient to capture both the global dispersion and its detailed structure.

\section{Phonon modes and electric-field response}

In this section, we identify the phonon modes that couple to an external electric field and quantify their lattice response.
\begin{figure}[t]
  \centering
  \includegraphics[width=\columnwidth]{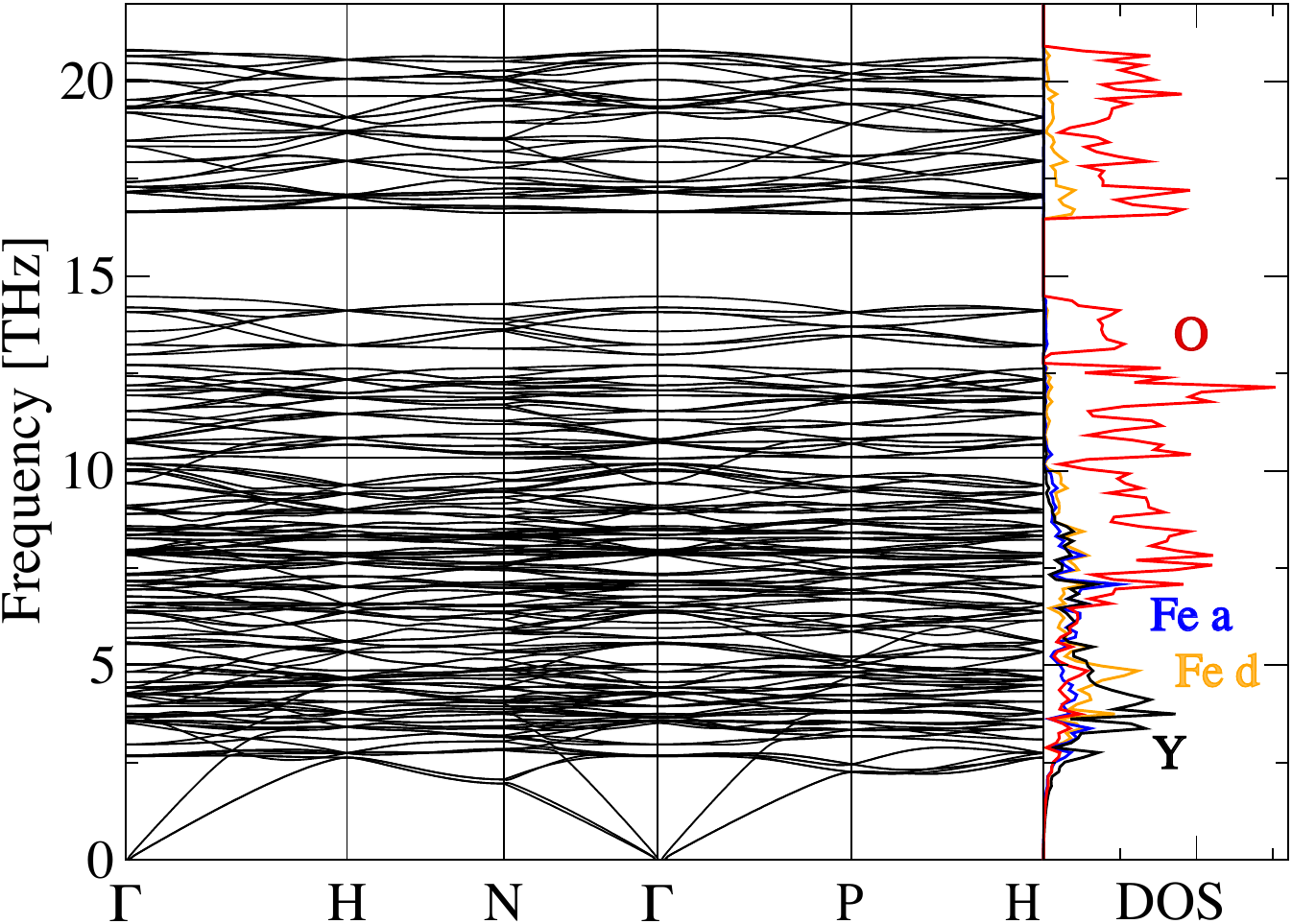}%{banddos.pdf}
  \caption{
Phonon band structure calculated along the high-symmetry path
$\Gamma$--H--N--$\Gamma$--P (left panel), together with the corresponding projected phonon density of states (DOS) (right panel, in arbitrary units).
The projected DOS is resolved into atomic contributions from Y (black), Fe at the $d$ sites (orange), Fe at the $a$ sites (blue), and O (red).
  }
  \label{fig:phonon_band}
\end{figure}
Figure~\ref{fig:phonon_band} shows the calculated phonon band structure of YIG.
The acoustic modes emerge from zero frequency at the $\Gamma$ point, while a large number of optical modes extend up to approximately 20~THz, reflecting the large unit cell.
Small imaginary frequency ($-0.06 < \omega < 0$ THz) at the $\Gamma$ point is a numerical artifact due to imperfect enforcement of the acoustic sum rule. 
The phonon and magnon energy scales are comparable, enabling efficient coupling between lattice and spin degrees of freedom. 
The projected phonon density of states indicates that low-frequency modes are dominated by Y vibrations, intermediate-frequency modes involve significant contributions from Fe atoms at both $a$ and $d$ sites, and high-frequency modes are primarily associated with oxygen vibrations reflecting the atomic masses of Y, Fe, and O.\cite{Wang2024} 

Among these modes, infrared-active optical phonons at the Brillouin-zone center are of particular importance.
Although YIG is centrosymmetric and does not exhibit a static polarization, these modes carry finite mode effective charges and therefore may couple directly to an external electric field.
The lattice response to an electric field can be described in terms of normal modes as
\begin{equation}
Q_{m,\alpha}^{(\mathrm{resp})} = \frac{Z_{m,\alpha}}{\omega_m^2} E_\alpha,
\end{equation}
where $Q_{m,\alpha}^{(\mathrm{resp})}$ is the displacement amplitude of mode $m$ along direction $\alpha$, $\omega_m$ is the phonon frequency, and $Z_{m,\alpha}$ is the mode-resolved effective charge defined as
\begin{equation}
Z_{m,\alpha} = \sum_s \sum_\beta Z^{*}_{s,\alpha\beta} \, e_{s,m}^{\beta}.
\end{equation}
Here, $Z^{*}_{s,\alpha\beta}$ is the Born effective charge tensor and $e_{s,m}^{\beta}$ is the phonon eigenvector.
This relation shows that the electric-field-induced displacement is enhanced for modes with large effective charge and low frequency.

\begin{figure*}[t]
  \centering
  \includegraphics[width=1.0\linewidth]{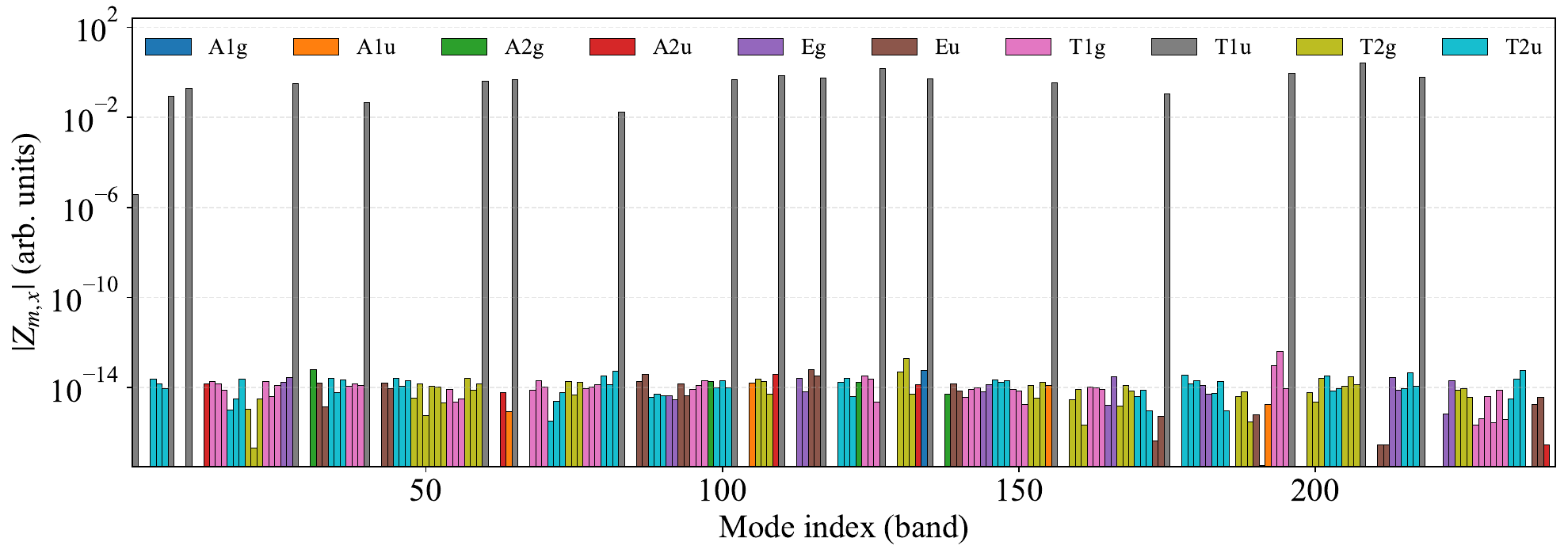} %{Qm_x_all_modes.pdf}
\caption{
Phonon-mode-resolved effective charge in \YIG.
%The response is characterized by the mode effective charge $Z_{m,x}$ for each phonon mode $m$. % defined as $Z_{m,x} = Q^{(\mathrm{resp})}_{m,x}\,\omega_m^2$.
A pronounced enhancement is observed for infrared-active $T_{1u}$ modes, indicating their dominant role in the electric-field response.
}
  \label{fig:Qm}
\end{figure*}

Figure~\ref{fig:Qm} shows the mode-resolved effective charge.
Only the infrared-active $T_{1u}$ modes exhibit sizable values of $Z_{m,\alpha}$, while all other modes are negligible.
Among these, several modes show particularly large values, corresponding to relative displacements of Fe and O ions with opposite effective charge moving in opposite directions, and are therefore efficiently excited by an external electric field.
Their impact on magnetic interactions is analyzed in the following section.

\section{Effect of Frozen Phonons on Magnetic Interactions}

We examine how the phonon modes affect the magnetic interactions in YIG.
In particular, the infrared-active $T_{1u}$ modes, which dominate the electric-field response, induce ionic displacements that modify the nearest-neighbor exchange interaction $J_{ad}$ via the exchange-striction mechanism.
To quantify this effect, ionic displacements following the phonon eigenvectors are introduced into the crystal structure (see Appendix~\ref{app:phonon_alignment}) using a frozen-phonon approach, and the resulting changes in the magnetic interactions are evaluated in a mode-resolved manner.  

\begin{figure}[t]
  \centering
  \includegraphics[width=\linewidth]{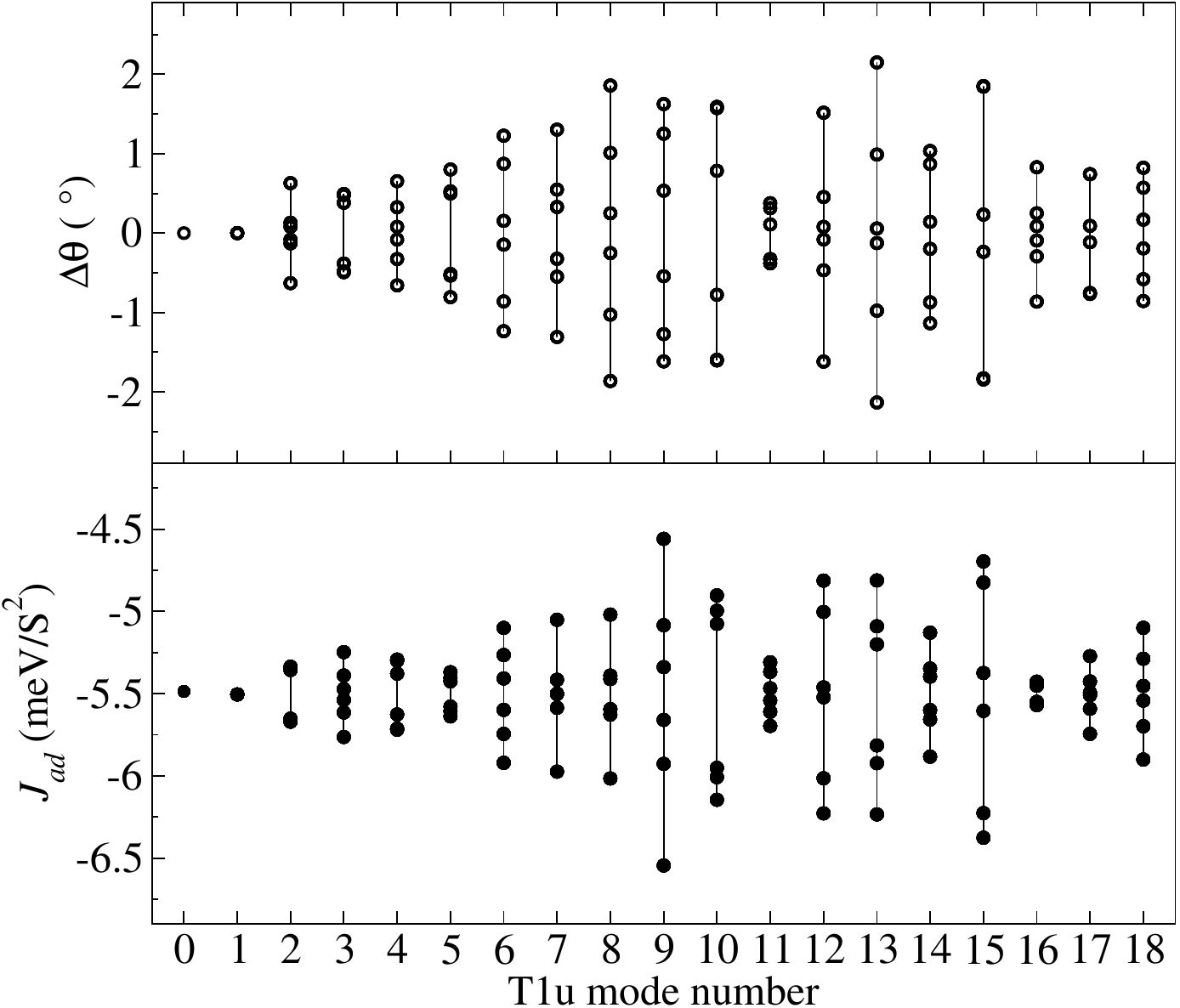} %{angle_J1.pdf}
  \caption{
Correlation between the phonon-induced variation of the Fe$_{a}$--O--Fe$_{d}$ bond angle $\Delta\theta$
(measured from the equilibrium structure)
and $J_{ad}$ induced by $T_{1u}$ phonon modes.
  }
  \label{fig:scatter}
\end{figure}

Figure~\ref{fig:scatter} shows the variation of $J_{ad}$ under phonon-induced ionic displacements and the corresponding bond-angle changes. 
A similar trend is observed between the bond-angle variation and $J_{ad}$.
Only minor changes are found for the lowest modes (mode~1--5), which lie in the low-frequency range below $5$~THz and are dominated by Y vibrations, whereas the variation becomes pronounced for higher-index modes (mode~6 and above), located above $5$~THz and involving significant contributions from both Fe and O atoms.
Among these, mode~9 produces the largest modulation of $J_{ad}$, with a frequency of $7.91$~THz at the $\Gamma$ point, lying in the range where both Fe and O atoms contribute to the phonon density of states (see Fig.~\ref{fig:phonon_band} and Tbl.\ref{tab:T1u_modes}).

\begin{figure}[t]
  \centering
  \includegraphics[width=0.7\linewidth]{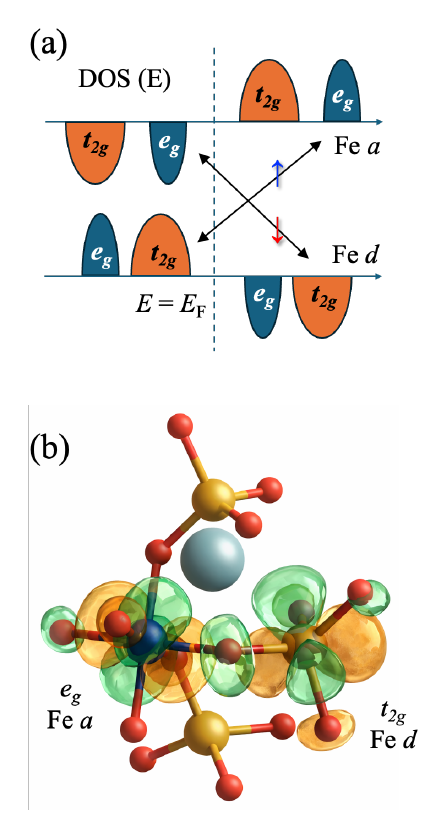} %{orbitallevel.pdf}
\caption{
(a) Schematic picture of superexchange between Fe$_{a}$ (octahedral) and Fe$_{d}$ (tetrahedral) sites. Black arrows represent virtual hopping processes, while the short arrows (blue and red) indicate majority- and minority-spin states, respectively. 
(b) Wannier functions associated with the dominant hopping processes contributing to $J_{ad}$.
}
  \label{fig:wan_pddp}
\end{figure}

\begin{figure}[t]
\centering
\includegraphics[width=0.8\linewidth]{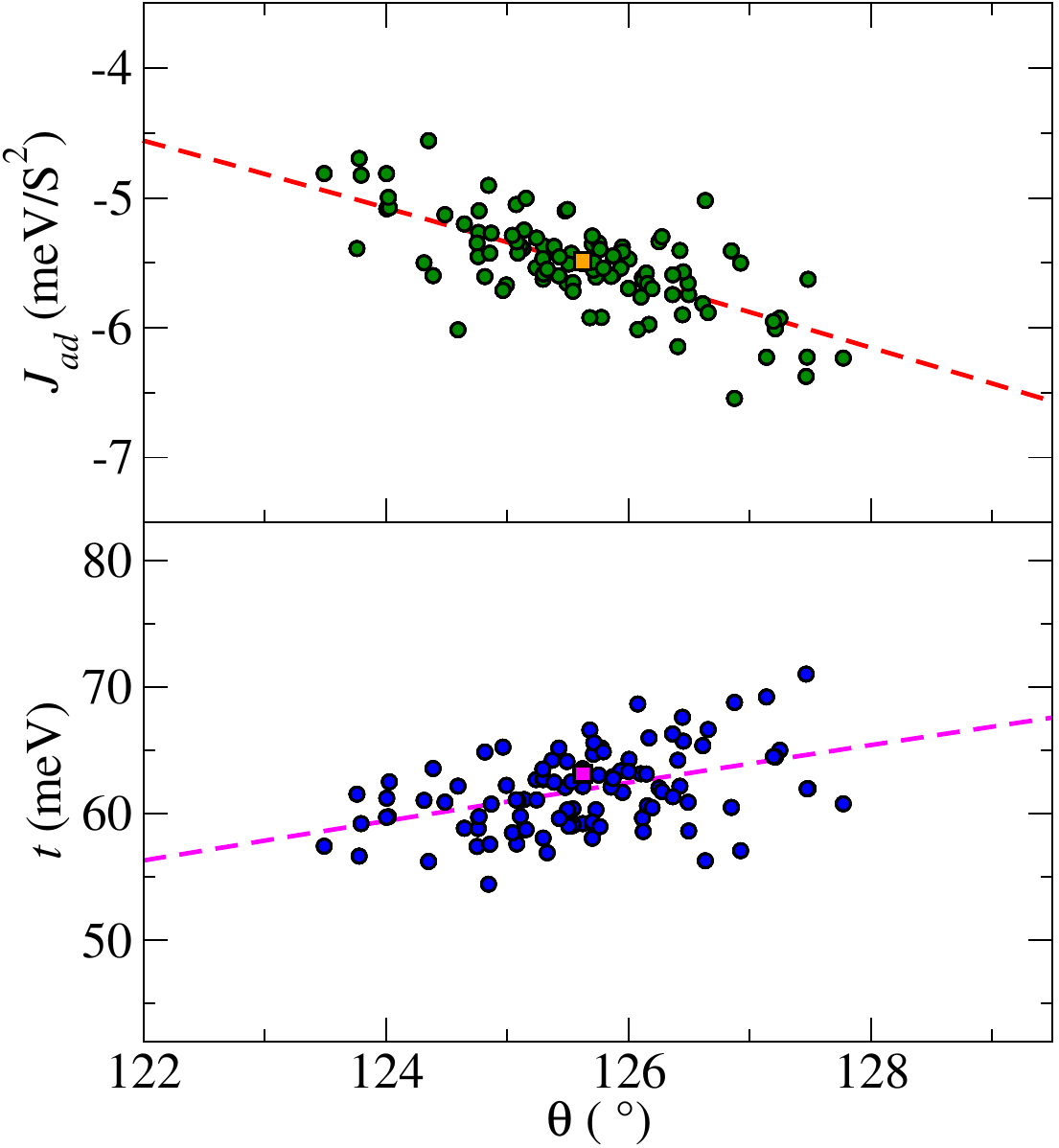} %{angle_J1andTdd.pdf}
\caption{
Dependence of (top) the nearest-neighbor exchange interaction $J_{ad}$ and (bottom) the dominant $d$--$d$ hopping amplitude $t_{dd}$ on the Fe$_a$--O--Fe$_d$ bond angle $\theta$ for the $T_{1u,x}$ phonon modes (mode~1--18).
All 21 Fe$_a$--O--Fe$_d$ bond angles in the unit cell are equivalent in the undistorted structure, but become inequivalent under the phonon-induced ionic displacements, and all resulting bond angles are included in the plot. 
Dashed lines indicate approximate $\cos^2\theta$ (top) and $\cos\theta$ (bottom) dependencies.
The open square denotes the undistorted reference structure.
}
\label{fig:angle_J1andTdd}
\end{figure}

To elucidate the microscopic origin of the dominant nearest-neighbor exchange interaction $J_{ad}$, we analyze the Wannier functions associated with the large $d$--$d$ hopping amplitudes; in the present Wannier construction, the Fe $d$ orbitals already incorporate hybridization with the surrounding O $p$ states, so that the effective $d$--$d$ hopping implicitly contains oxygen-mediated processes. 
Multiple hopping channels exist between Fe$_{a}$ and Fe$_{d}$ sites, with the dominant contribution arising from $e_g$--$t_{2g}$ hopping as shown in Fig.~\ref{fig:wan_pddp}. 
The exchange interaction is mediated by virtual hopping processes between occupied and unoccupied orbitals, namely between occupied $e_g$ and unoccupied $t_{2g}$ states for the majority-spin channel, as well as between occupied $t_{2g}$ and unoccupied $e_g$ states for the minority-spin channel~\cite{Anderson1959,Nguyen2021}.
As the Fe$_a$--O--Fe$_d$ bond angle changes, the orbital-dependent hopping processes are modified, and the resulting superexchange interaction follows the Goodenough--Kanamori--Anderson rules~\cite{Goodenough1955,Kanamori1959,Anderson1959}. 
The antiferromagnetic contribution to the exchange interaction is then approximated as
\begin{equation}
J_{ad}^{\mathrm{AF}}(\theta) \simeq \frac{4|t_{dd}(\theta)|^2}{U_{\mathrm{eff}}},
\end{equation}
where $U_{\mathrm{eff}}$ denotes the effective excitation energy associated with the virtual hopping process.
The angular dependence is primarily governed by the $(d$--$p)\pi$ overlap and can be captured by a minimal parameterization $t_{dd}(\theta) = t_{dd}^{(0)} |\cos\theta|^n$, leading to $J_{ad}(\theta) \propto |\cos\theta|^{2n}$.\cite{Anderson1959,Goodenough1955,Kanamori1959,Slater1954} 
As shown in Fig.~\ref{fig:angle_J1andTdd}, both the exchange interaction and the hopping amplitude depend on the Fe$_a$--O--Fe$_d$ bond angle, showing roughly $\cos^2\theta$ and $\cos\theta$ behaviors, respectively, in line with the Goodenough--Kanamori--Anderson rules. 

Optical phonons induce substantial variations in the exchange interaction $J_{ad}$, with the magnitude of the modulation depending strongly on the phonon mode.
Modes involving Fe and O displacements produce the largest effects.
Although magnon--phonon coupling arising from spin--orbit interaction is expected to be weak in YIG, the exchange-striction mechanism is sizable and may contribute to magnon--phonon coupling.
A minimal model describing the resulting modifications of the magnon dispersion is discussed in Appendix~\ref{app:mp_model}.

\section{Conclusion}

We have investigated the interplay between lattice and spin degrees of freedom in yttrium iron garnet from a first-principles perspective, combining electronic structure calculations, phonon analysis, and a Wannier-based evaluation of magnetic exchange interactions.
The resulting exchange parameters reproduce the magnon dispersion and confirm that the nearest-neighbor interaction $J_{ad}$ sets the dominant energy scale.

A mode-resolved frozen-phonon approach has been used to examine how lattice displacements modify the exchange interactions.
The modulation of $J_{ad}$ depends strongly on the phonon mode, with optical modes involving Fe and O displacements producing the largest variations.
Although magnon--phonon coupling arising from spin--orbit interaction is expected to be weak in YIG, the exchange-striction mechanism can be sizable and may contribute to the coupling between magnons and phonons.
Such effects may be accessible experimentally through the excitation of infrared-active phonons, for example by THz pulses, suggesting that phonon excitation can modify magnon properties and thereby influence spin-wave dynamics.

\begin{acknowledgments}
We are grateful to Hitoshi Tabata and Munetoshi Seki for the fruitful discussions. 
This work was supported  by JST-CREST (Grant No. JPMJCR22O2) and by
Institute for Open and Transdisciplinary Research Initiatives (OTRI), tthe University of Osaka.
The computation in this work has been done using the facilities of the Supercomputer Center, the Institute for Solid State Physics, the University of Tokyo and the supercomputer ``Flow'' at the Information Technology Center,
Nagoya University. The crystallographic figure was generated using the VESTA program~\cite{Momma2011}.
\end{acknowledgments}
%============================================================
\appendix

% ============================================================

\section{Linear spin-wave theory for ferrimagnetic YIG}
\label{app:lswt}

The magnon spectrum of YIG was computed by the linear spin-wave theory (LSWT) formulation.\cite{Holstein1940, Anderson1952} 
%YIG is a ferrimagnet consisting of two magnetic sublattices: octahedral $a$ sites and tetrahedral $d$ sites.
In the primitive body-centered cubic cell of  YIG, there are $N_a=8$ $a$-site Fe ions and $N_d=12$ $d$-site Fe ions, giving a total of $N=20$ spins.
The classical ground state is collinear, with spins on the two sublattices aligned antiparallel.

We start from a shell-resolved Heisenberg Hamiltonian
\begin{align}
\mathcal{H}
= -&\sum_{\langle i,j\rangle \in ad} J_{ad}^{(n)} \,\mathbf{S}_i \cdot \mathbf{S}_j \nonumber \\
  &- \sum_{\langle i,j\rangle \in dd} J_{dd}^{(n)} \,\mathbf{S}_i \cdot \mathbf{S}_j
   - \sum_{\langle i,j\rangle \in aa} J_{aa}^{(n)} \,\mathbf{S}_i \cdot \mathbf{S}_j ,
\label{eq:H_app}
\end{align}
where $n$ labels neighbor shells identified geometrically.
A negative $J$ favors antiparallel alignment.

Local spin quantization axes are chosen such that spins on the $a$ sublattice point along $+z$ and those on the $d$ sublattice along $-z$.
The spin operators are expanded using the Holstein--Primakoff transformation\cite{Holstein1940},
\begin{align}
S_i^z &= \sigma_i \left(S - a_i^\dagger a_i\right), \qquad
\sigma_i =
\begin{cases}
+1 & (a\text{ site}) \\
-1 & (d\text{ site})
\end{cases}, \\
S_i^+ &\simeq \sqrt{2S}\, a_i, \qquad
S_i^- \simeq \sqrt{2S}\, a_i^\dagger ,
\end{align}
keeping terms up to quadratic order in bosonic operators.

After Fourier transformation, the quadratic Hamiltonian can be written in Nambu form,
\begin{equation}
\mathcal{H}
= \frac{1}{2}\sum_{\mathbf{k}}
\Psi_{\mathbf{k}}^\dagger
\begin{pmatrix}
A_{\mathbf{k}} & B_{\mathbf{k}} \\
B_{\mathbf{k}}^\dagger & A_{-\mathbf{k}}^{\mathsf{T}}
\end{pmatrix}
\Psi_{\mathbf{k}}
+ \mathrm{const.},
\label{eq:Hk_app}
\end{equation}
where
\[
\Psi_{\mathbf{k}} =
(a_{1\mathbf{k}}, \ldots, a_{20\mathbf{k}} \mid
a_{1,-\mathbf{k}}^\dagger, \ldots, a_{20,-\mathbf{k}}^\dagger)^{\mathsf{T}} .
\]

The matrices $A_{\mathbf{k}}$ and $B_{\mathbf{k}}$ are $20\times20$ matrices.
$A_{\mathbf{k}}$ contains number-conserving hopping and onsite terms, while $B_{\mathbf{k}}$ contains anomalous terms that mix creation and annihilation operators.
The presence of $B_{\mathbf{k}}$ is a direct consequence of the ferrimagnetic ground state and is essential for obtaining the correct magnon spectrum.

% ============================================================
%\section{Bosonic Bogoliubov--de Gennes diagonalization}
%\label{app:bdg}

The quadratic bosonic Hamiltonian\cite{Colpa1978} in Eq.~\eqref{eq:Hk_app} is diagonalized using the bosonic Bogoliubov--de Gennes (BdG) formalism.
The eigenvalue problem reads
\begin{equation}
\eta \mathcal{H}_{\mathbf{k}} \mathbf{v}_{\mathbf{k},n}
= \omega_{\mathbf{k},n} \mathbf{v}_{\mathbf{k},n},
\label{eq:bdg_app}
\end{equation}
with the paraunitary metric
\begin{equation}
\eta = \mathrm{diag}(\underbrace{1,\ldots,1}_{20},\underbrace{-1,\ldots,-1}_{20}).
\end{equation}

The normalization condition is
\begin{equation}
\mathbf{v}_{\mathbf{k},m}^\dagger \eta \mathbf{v}_{\mathbf{k},n}
= \delta_{mn}.
\end{equation}
Eigenvalues appear in $\pm\omega$ pairs, and the physical magnon spectrum consists of the 20 positive eigenvalues $\omega_{\mathbf{k},n}>0$.

A naive diagonalization of the $20\times20$ matrix $A_{\mathbf{k}}$ alone neglects the anomalous terms in $B_{\mathbf{k}}$ and produces an unphysical spectrum in which the high-energy branches are mirror images of the low-energy ones.
Only the full $40\times40$ bosonic BdG problem yields the correct separation into acoustic and optical magnon branches.
This is a general feature of ferrimagnetic and antiferromagnetic systems.

To track magnon bands along high-symmetry paths, we connect eigenmodes at neighboring $\mathbf{k}$ points by maximizing the overlap
\begin{equation}
\mathcal{O}_{mn}(\mathbf{k}\rightarrow\mathbf{k}')
= \left| \mathbf{v}_{\mathbf{k},m}^\dagger
\eta \mathbf{v}_{\mathbf{k}',n} \right|.
\end{equation}
This procedure ensures smooth band connectivity and avoids artificial band crossings.
%

%\section{Remarks on $S(S+1)$ factors and classical mapping}
%\label{app:spin_length}
The exchange interactions $J_{ij}$ are extracted by mapping total energies of collinear spin configurations onto a classical Heisenberg model.
As a result, the fitted exchange parameters may implicitly include factors of the spin length $S$.
In LSWT, the overall magnon energy scale depends on the combination of $J_{ij}$ and $S$.
In the present work, we use $S=5/2$ for Fe$^{3+}$ ions and apply a consistent convention throughout all calculations, enabling direct comparison between different $U$ values and with experimental data.

% ============================================================
%\section{Brillouin zone and high-symmetry paths}
%\label{app:bz}

%The primitive cell of YIG is body-centered cubic.
%Magnon dispersions are computed along the conventional high-symmetry path $N$--$\Gamma$--$H$ of the bcc Brillouin zone.
%The $\Gamma$ point corresponds to the Goldstone mode associated with global spin rotation, and the acoustic magnon branch vanishes at $\Gamma$ within numerical accuracy.
%
%Zone-boundary degeneracies at $N$ and $H$ arise from the nonsymmorphic symmetry of the garnet structure and are correctly reproduced by the present LSWT implementation.

% ============================================================
\section{Post-processing and definition of $T_{1u}$ phonon modes}
\label{app:phonon_alignment}

Phonon properties were computed using the finite-displacement method as implemented in \textsc{phonopy}.
In this work, we focus on zone-center ($\Gamma$-point) optical phonons, in particular the infrared-active $T_{1u}$ modes, which couple directly to an external electric field.
To evaluate mode-resolved quantities such as bond-angle changes and exchange interactions, a well-defined representation of the phonon eigenvectors is required, and a post-processing procedure is applied to the raw eigenvectors obtained from \textsc{phonopy}.

\begin{table}[t]
\centering
\caption{$T_{1u}$ phonon frequencies at the $\Gamma$ point. The mode index is assigned in ascending order of frequency.}
\label{tab:T1u_modes}
\begin{tabular}{cc|cc}
\hline
Mode & Frequency (THz) & Mode & Frequency (THz) \\
\hline
mode01 & $\sim 0$  & mode10 & 8.26 \\
mode02 & 2.69      & mode11 & 8.47 \\
mode03 & 2.97      & mode12 & 9.03 \\
mode04 & 3.95      & mode13 & 9.68 \\
mode05 & 4.43      & mode14 & 10.81 \\
mode06 & 5.70      & mode15 & 12.43 \\
mode07 & 6.36      & mode16 & 16.66 \\
mode08 & 7.05      & mode17 & 17.42 \\
mode09 & 7.91      & mode18 & 19.20 \\
\hline
\end{tabular}
\end{table}

At the Brillouin-zone center, phonon modes in cubic crystals appear as degenerate multiplets due to the high point-group symmetry.
In YIG, the $T_{1u}$ modes are triply degenerate and correspond to Cartesian polarizations along $x$, $y$, and $z$.
The corresponding frequencies are listed in Table~\ref{tab:T1u_modes}.

Numerically, phonon codes return an arbitrary orthonormal basis within each degenerate subspace.
To obtain a physically meaningful representation, the eigenvectors are rotated into a Cartesian basis by constructing an orthogonal transformation under a mass-weighted inner product.
This is achieved by projecting the phonon eigenvectors onto reference displacement patterns along the Cartesian directions and maximizing their overlap, thereby fixing the gauge of the degenerate subspace.

The atomic displacements used in the frozen-phonon calculations are constructed from the rotated eigenvectors as
\begin{equation}
\mathbf{u}_{i}^{(\nu)} = A \, \frac{\mathbf{e}_{i}^{(\nu)}}{\sqrt{M_i}},
\end{equation}
where $\mathbf{e}_{i}^{(\nu)}$ is the Cartesian phonon eigenvector of atom $i$, $M_i$ is the atomic mass, and $A$ is a global amplitude factor.
The distorted structures are generated as
\begin{equation}
\mathbf{R}_i^{\pm} = \mathbf{R}_i^{(0)} \pm \mathbf{u}_{i}^{(\nu)}.
\end{equation}
In the present calculations, a fixed amplitude $A=1.0$ is used, and the resulting maximum displacement
\begin{equation}
u_{\max} = \max_i \left| \mathbf{u}_{i}^{(\nu)} \right|
\end{equation}
is typically in the range of $10^{-2}$--$10^{-1}$~\AA, confirming that the distortions remain within the linear-response regime.

Finally, the mode effective charge is evaluated as
\begin{equation}
\widetilde{\mathbf{Z}}_m
= \sum_s \mathbf{Z}_s^* \cdot \mathbf{e}_{s,m},
\end{equation}
where $\mathbf{Z}_s^*$ is the Born effective charge tensor.
Only the $T_{1u}$ modes have nonzero $\widetilde{\mathbf{Z}}_m$ and therefore couple linearly to an external electric field.

% ============================================================

% ============================================================

\section{Minimal magnon--phonon hybridization model}
\label{app:mp_model}

\begin{figure}[t]
  \centering
  \includegraphics[width=0.9\linewidth]{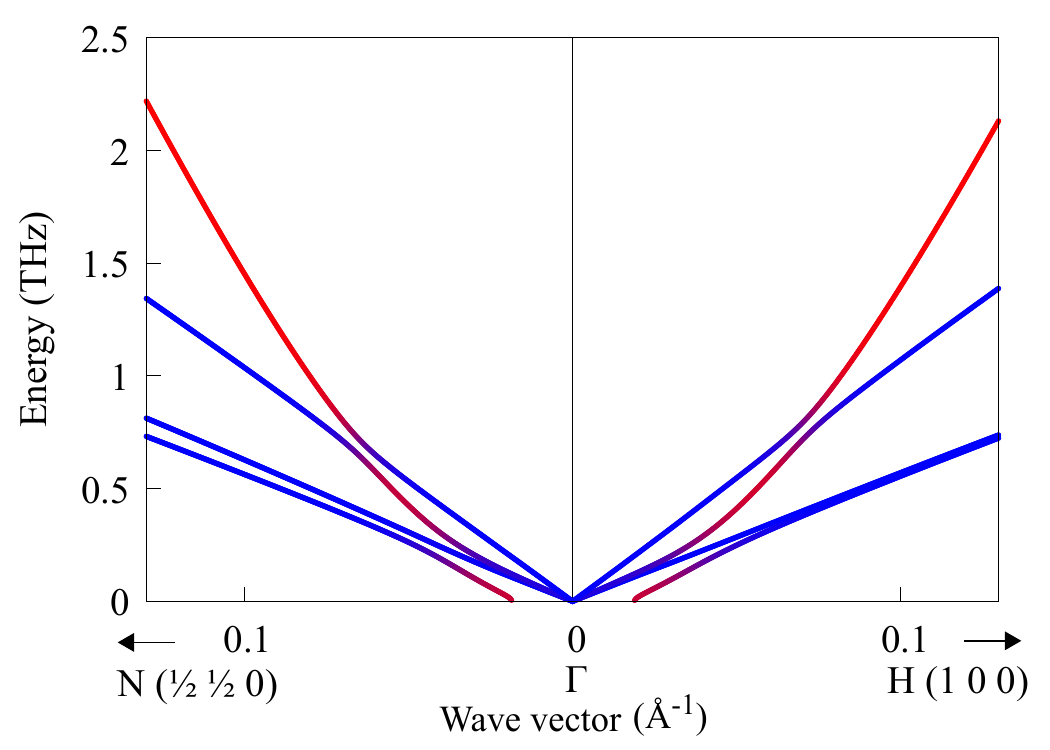} %{hybrid_zoom.pdf}
  \caption{
Hybridized magnon--phonon band structure obtained from the minimal model.
The color indicates the magnon weight.
A zoom around the $\Gamma$ point along N--$\Gamma$--H path is shown.
}
  \label{fig:mpband}
\end{figure}

To describe the avoided crossings between magnons and transverse acoustic (TA) phonons, we consider a minimal effective model.\cite{Flebus2017PRB}

We take one magnon mode and three acoustic phonon modes (one longitudinal and two transverse) at each $\mathbf{k}$ point, and construct the effective dynamical matrix %in $\omega^2$ space 
as
\begin{equation}
D_{\mathrm{eff}}(\mathbf{k}) =
\begin{pmatrix}
\omega_{\mathrm{m}}^2(\mathbf{k}) &
g^2(\mathbf{k}) &
g^2(\mathbf{k}) &
g^2(\mathbf{k}) \\
g^2(\mathbf{k}) &
\omega_{\mathrm{ph},1}^2(\mathbf{k}) &
0 &
0 \\
g^2(\mathbf{k}) &
0 &
\omega_{\mathrm{ph},2}^2(\mathbf{k}) &
0 \\
g^2(\mathbf{k}) &
0 &
0 &
\omega_{\mathrm{ph},3}^2(\mathbf{k})
\end{pmatrix}.
\end{equation}

Here $\omega_{\mathrm{m}}(\mathbf{k})$ is the magnon dispersion obtained from spin-wave calculations based on first-principles exchange parameters, and $\omega_{\mathrm{ph},i}(\mathbf{k})$ are the phonon dispersions from first-principles calculations.
The model coupling is taken as
\begin{equation}
g(\mathbf{k}) = g_0 |\mathbf{k}-\mathbf{k}_0|.
\end{equation}
The hybridized energies are obtained from the eigenvalues $\lambda_i(\mathbf{k})$ of $D_{\mathrm{eff}}(\mathbf{k})$ as $\omega_i(\mathbf{k})=\sqrt{\lambda_i(\mathbf{k})}$.

As shown in Fig.~\ref{fig:mpband}, the coupling produces avoided crossings between the magnon and TA phonon branches, and a strong mixing of modes near the $\Gamma$ point where multiple branches approach each other.

Since both the magnon and phonon dispersions are obtained from first-principles calculations, the crossing positions are quantitatively reliable, while the coupling strength is introduced phenomenologically.
Experimental verification of these avoided crossings would provide a direct test of the exchange-striction mechanism discussed in this work.

\bibliographystyle{apsrev4-2}
\bibliography{yig_refs}

\end{document}